\documentclass[reprint,aps,prc,showpacs,floatfix]{revtex4-1}
\usepackage{graphics,epsfig,graphicx}
\usepackage{amsbsy}
\usepackage{amsmath}
\usepackage{bm}
\usepackage{amsfonts}
\usepackage{cancel}
\usepackage{multirow}
\usepackage{slashed}
\begin{document}

\title{Implications of Efimov physics for the description of three and 
   four nucleons in chiral effective field theory}

\author{A. Kievsky and M. Viviani} 
\affiliation{Istituto Nazionale di Fisica Nucleare, Largo Pontecorvo 3, 56100 Pisa, Italy}
\author{M. Gattobigio}
\affiliation{Universit\'e de Nice-Sophia Antipolis, Institut Non-Lin\'eaire de
Nice,  CNRS, 1361 route des Lucioles, 06560 Valbonne, France }
\author{L. Girlanda}
\affiliation{Department of Mathematics and Physics, University of
Salento, 73100 Lecce, Italy and INFN-Lecce, Lecce, Italy}

\begin{abstract}
In chiral effective field theory the leading order (LO) nucleon-nucleon potential includes two contact terms,
in the two spin channels $S=0,1$, and the one-pion-exchange potential. 
When the pion degrees of freedom are integrated out, as in the pionless
effective field theory, the LO potential includes two contact terms only.
In the three-nucleon system, the pionless theory includes a three-nucleon contact
term interaction at LO whereas the chiral effective theory does not. Accordingly arbitrary differences could
be observed in the LO description of three- and four-nucleon binding energies. We analyze the two
theories at LO and conclude that a three-nucleon contact term is necessary at this order in both
theories. 
In turn this implies that subleading three-nucleon contact terms should be promoted to lower orders.
Furthermore this analysis shows that one single low energy constant might be sufficient to
explain the large values of the singlet and triplet scattering lengths.
\end{abstract}
\maketitle

\section{Introduction} 
Strong efforts have been done in recent years to determine the nuclear interaction
from first principles. The starting point is an effective Lagrangian in terms of the
relevant degrees of freedom, nucleons and pions, maintaining the symmetries of the strong interaction, 
and a perturbative framework named chiral perturbation theory (ChPT)
(see Refs.~\cite{epelbaum06,epelbaum09} and references therein). 
This setting  allowed to construct
two and three-body interactions up to next-to-next-to-next-to-next leading order (N4LO).
Though potentials up to N5LO have been studied~\cite{entem15,epelbaum15},
at present most of the calculations in finite nuclei and infinite nuclear matter have been done
with the nucleon-nucleon (NN) interaction up to N3LO and the three-nucleon 
interaction up to N2LO. At LO the chiral perturbative expansion includes two contact
terms, characterized by the low energy constants (LECs) $C_S$ and $C_T$ governing the (spin) 
singlet and triplet $s$-waves, and the (regularized) one-pion-exchange potential (OPEP).
At very low energies the pion degrees of freedom can be integrated out resulting in what is
called pionless effective theory~\cite{bedaque02}. At LO this theory includes only two contact
terms without the inclusion of the pionic tail. In the low energy regime the chiral and pionless  
effective theories should be equivalent. In fact both theories have two LECs to be determined by two observables, 
such as e.g. the singlet and triplet scattering lengths, implying an equivalent description
of the low energy $s$-wave scattering. Since the OPEP tail includes the tensor interaction,
small differences between the two theories arise from those
observables depending on the $d$-state component of the deuteron as for example the quadrupole moment.
However these quantities are small and can be considered to be inside the theoretical error produced by
the truncation of the perturbative expansion at LO. Accordingly
both theories give an equivalent description of the two nucleon system at LO. 

The extension to the three-nucleon sector produces a drastic difference between
the two theories and this is the motivation of the present investigation. In chiral perturbation theory
there is no new term at LO when the three-nucleon system is considered~\cite{epelbaum02}. Accordingly the LO NN potential
used in the two-body system has to be used to describe the three-body system without the
inclusion of any new LEC. Conversely, and it is well known from Efimov physics, the pionless
theory includes a contact three-body force at LO with a new LEC that can be used to determine the energy
of the three-nucleon bound state~\cite{bedaque1999}. 
Assuming that there are no
other many-body forces at LO, systems with $A\ge3$ are described in the LO pionless theory with 
a sum of a two- plus a three-body term whereas the LO chiral potential consists only 
in a two-body term. Since the new LEC can be used to fix the triton binding energy, we expect a better 
description of the $^3$H, $^3$He and $^4$He binding energies
using the LO pionless potential than using the LO chiral one. 
As a result, the pionless expansion can be expected to converge better than the chiral one. In principle there is no 
contradiction: ChPT, being a more microscopic theory, yields a more economic description, with less unknowns than 
the pionless theory; differences in the convergence pattern are also expected, as the perturbative expansions are 
different in the two theories (e.g. the expansion parameters are different). Nevertheless we consider this situation 
as undesirable from a practical point of view, and think that this difference in the description of the $s$-wave nuclei 
at LO needs a deeper analysis.

At the base of the difference between the two theories is the fact that the singlet and triplet
scattering lengths $a_0\approx -24\;$fm and $a_1\approx 5.4\;$fm are large with respect to
the range of the nuclear interaction $r_N\approx 1.4\;$fm. A direct consequence is the shallow
characteristic of the deuteron binding energy $E_d\approx \hbar^2/m a_1^2$ with corrections in
powers of the small quantity $r^1_{eff}/a_1$, with $r^1_{eff}\approx r_N$ the effective range in the triplet
state. The existence of large scattering lengths puts the two-, three- and four-nucleon
systems inside a window in which the concepts of the Efimov physics can be applied (see
Refs.~\cite{kievsky2016,koenig2016} for a recent discussion). This
physics describes systems close to the unitary limit, $a_0,a_1\rightarrow\infty$, using a zero-range
theory. In this limit the three-body system is characterized by an infinite series of excited states, 
called Efimov effect~\cite{efimov1,efimov2}.
Close to the unitary limit there is no length scale and the three-body contact term
is needed to stabilize the system against variations of the cutoff introduced at the two-body level
to regularize the theory. It could be thought that since the OPEP tail introduces a length scale
there is no need for a three-body counter term to stabilize the theory even if the system is close to
the unitary limit. However, as we are dealing with shallow states, the inclusion of the potential tail
is in competition with the zero-range description suggested by the large values of the
scattering lengths. Accordingly there is a noticeable sensitivity to variations of the
cutoff, though smaller compared to the pionless case, not completely dampened by the inclusion
of the OPEP tail. 

\section{Numerical analysis} 
We perform a numerical analysis of both
theories at LO in order to assess the necessity of including a three-body force at LO
in both theories. We start analyzing the two body sector. The LO effective Hamiltonian 
for two nucleons can be put in the form
\begin{equation}
H_{LO}= -\frac{\hbar^2\nabla^2}{m_N} + V_{sr} +V_\pi
\end{equation}
where $V_{sr}$ is the (regularized) short-range contact interaction and $V_\pi$ is the OPEP. The short-range
interaction, considered to act in $s$-waves, has a spin dependence and can be written as
\begin{equation}
V_{sr}=C_S V_c + C_T V_\sigma {\bm \sigma}_1\cdot{\bm \sigma}_2 \;\; .
\end{equation}
Using a local gaussian regulator, the two potentials $V_c$ and $V_\sigma$ have
the following form
\begin{equation}
V_c=V_\sigma= V_0(r)= e^{-r^2/r_0^2}
\end{equation}
with $\Lambda_0=1/r_0$ the cutoff parameter. $V_\pi$ reads
\begin{equation}
V_\pi(r)= {\bm \tau}_1\cdot{\bm \tau}_2
\left[ {\bm \sigma}_1\cdot{\bm \sigma}_2 Y_\beta(r)+ S_{12} T_\beta(r)\right] 
\end{equation}
with the regularized factors ($x=m_\pi r$)
\begin{equation}
\begin{gathered}
Y_\beta(x)= \frac{g_A^2 m_\pi^3}{12\pi F_\pi^2}\frac{e^{-x}}{x}(1-e^{-r^2/\beta^2)} \\
T_\beta(x)= \frac{g_A^2 m_\pi^3}{12\pi F_\pi^2}\frac{e^{-x}}{x}(1+\frac{3}{x}+\frac{3}{x^2})(1-e^{-r^2/\beta^2})^2\;\; .
\end{gathered}
\end{equation}
Here $m_\pi=138.03\;$MeV is the average pion mass, $g_A=1.29$ is the nucleon coupling constant and
$F_\pi=2f_\pi=184.80\;$MeV is the pion decay amplitude. 

\begin{figure}
\includegraphics[height=\linewidth,angle=-90]{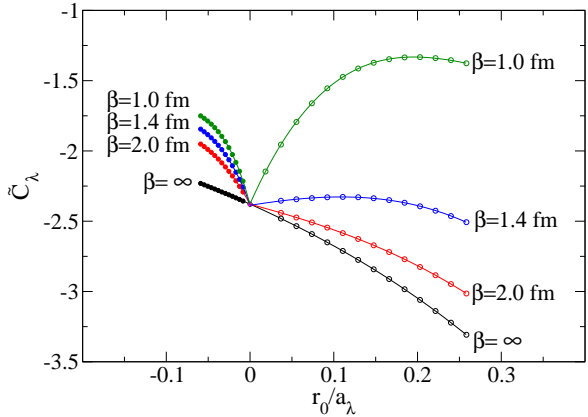}
\caption{Color online. RG trajectories of the coupling constant ${\widetilde C}_\lambda$ as a 
function of the inverse cutoff parameter $r_0$ in units of the singlet ($\lambda=0$, negative values)
and triplet ($\lambda=1$, positive values) scattering length $a_\lambda$ for different values of the 
regulator $\beta$.} \label{fig:fig1}
\end{figure}

Using $H_{LO}$, we solve the two-nucleon Schr\"odinger equation for different values of the cutoff 
$\Lambda_0$ 
and the regulator parameter $\beta$. For each case the two LECs $C_S$ and $C_T$ are selected in order to
reproduce the two scattering lengths, $a_0$ and $a_1$. It should be noticed that in the case 
$\beta\rightarrow \infty$ the chiral Hamiltonian $H_{LO}$ tends to the pionless one. 
In this case the coupling constants
$C_0=C_S-3C_T$ and $C_1=C_S+C_T$, corresponding to spin channels $S=0,1$ respectively, 
can be expanded in powers of the small parameters $r_0/a_\lambda$ as
\begin{equation}
{\widetilde C}_\lambda=\frac{\sqrt{\pi}}{2}\frac{mr_0^2}{\hbar^2}
C_\lambda=C_\infty\left(1+\alpha_1 \frac{r_0}{a_\lambda}+ \alpha_2 (\frac{r_0}{a_\lambda})^2 + \ldots\right)
\label{eq:rgt}
\end{equation}
with $\lambda=0,1$ and $C_\infty=2.379$~\cite{bazak2016,alvarez2016}. The pure number $2.379$ is
universal and gives the value of the coupling constant at the unitary limit. It also gives the
ratio $\sqrt{\pi}mr_0^2C_\lambda/(2\hbar^2)$ at the scaling limit,
$r_0\rightarrow 0$. Eq.~(\ref{eq:rgt}) maps the renormalization group (RG) trajectories as the interaction
approaches the scaling limit. In Fig.~\ref{fig:fig1} the different trajectories are shown as a fit of the results, 
given by the solid circles ($S=0$ state) and empty circles ($S=1$ state), 
as a function of $r_0/a_\lambda$ for different values of $\beta$. In all cases $C_\infty$ is 
independent of the spin state and of the regulator $\beta$ and represents the fixed
point to which all trajectories flow. Moreover, in the case of $\beta\to \infty$
the calculations of the spin $S=0,1$ states lie on a single trajectory.
The fact that all trajectories converge to the same fixed point can be seen as an indication
that both theories, pionless and chiral, need a three-body contact interaction to stabilize 
the three-body energy against variations in the cutoff $\Lambda_0$ (see below).

\begin{figure}
\includegraphics[height=\linewidth,angle=-90]{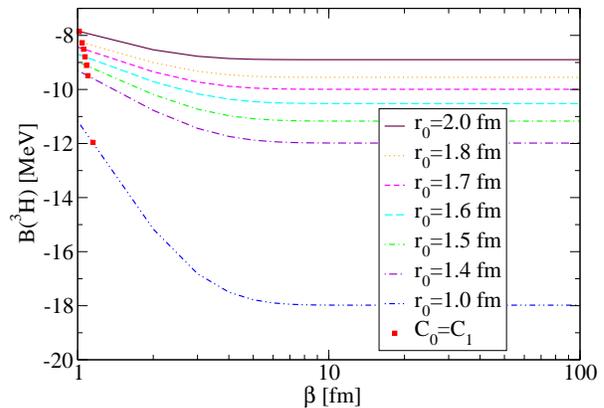}
\caption{Color online. The triton energy $B(^3{\rm H})$ as a function of the regulator $\beta$
for different values of the potential range $r_0$. The (red) squares indicate
the results obtained with $C_1=C_0$ corresponding to the $SU(4)_W$ symmetry.}
\label{fig:fig2}
\end{figure}

We extend our analysis to the three- and four-nucleon systems. The effective potential is 
\begin{equation}
\sum_{i\ne j}\left[V_{sr}(i,j)+V_\pi(i,j)\right] +\sum_{i\ne j \ne k} W(i,j,k)
\end{equation}
where we have added a (regularized) contact three-body term of the form
\begin{equation}
W(i,j,k)=W_0 e^{-r_{ij}^2/r_0^2} e^{-r_{ik}^2/r_0^2} \;\; .
\label{eq:w30}
\end{equation}
In first place we calculate the $^3$H energy, $B(^3{\rm H})$, considering only the two-body potential
terms for different values of $r_0$ and $\beta$. The results are shown in Fig.~\ref{fig:fig2} as a function of
the regulator of the OPEP $\beta$. The limit $\beta \rightarrow\infty$ corresponds to the pionless
theory whereas the lowest value, $\beta=1.0\;$fm, is well inside the region compatible with the formation of the
OPEP tail. 
We observe a noticeable spread in $B(^3{\rm H})$ for all the $r_0$ values considered, more pronounced
as $\beta \rightarrow\infty$. In the figure, the lowest result ($\approx -18\;$ MeV) corresponds 
to the shortest range considered as expected due to the Thomas collapse. For the lower values of $\beta$ the
spread in $B(^3{\rm H})$ is reduced, but not enough to judge the inclusion of the
OPEP tail sufficient for a stable description of $B(^3{\rm H})$ at LO. 
It should be noticed that extending the analysis to shorter
values of $r_0$ the spread increases more and more. So we can conclude that 
the reduction of the spread in the three-nucleon energy when the OPEP tail is included
is not significant, a three-body contact term is needed in both cases, pionless and chiral, 
to stabilize the results. 

In Fig.~\ref{fig:fig2} the (red) solid squares indicates those values of the
LECs verifying $C_1=C_0$ which implies that the short-range potentials includes only
the central term, $V_{sr}=C_S V_c$. This case corresponds to consider the Wigner spin-flavor symmetry 
$SU(4)_W$, which in turn is justified in the context of the large $N_c$ limit of QCD~\cite{hammer2000,vanasse2016}. 
Thus, for the 
corresponding values of the cutoff $\beta$, the difference between the singlet and triplet scattering 
length is entirely ascribed to the presence of the OPEP tail. Finer cutoff effects change the 
conclusion, but they may be considered to lie beyond the LO picture. It is reassuring to see that the 
fine-tuning which brings the NN system close to the unitary limit in both spin channels is only due to 
one single LEC, $C_S$.

\begin{figure}
\includegraphics[height=\linewidth,angle=-90]{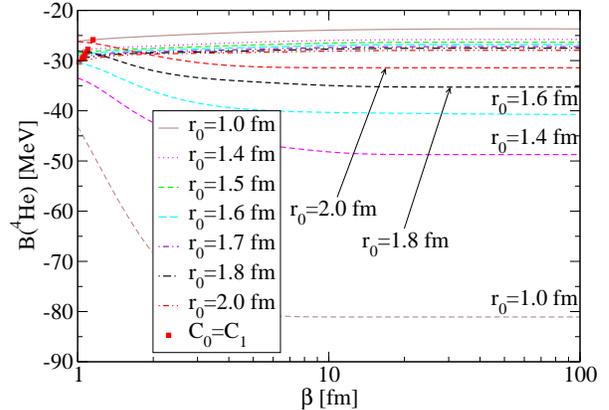}
\caption{Color online. The $B(^4{\rm He})$ binding energy as a function of the regulator $\beta$
for different values of the potential range $r_0$ with and without the three-body interaction. 
In the first case the $r_0$ values of the different curves are given in the box. The curves are 
very close to each other and are almost indistinguishable. In the second case the curves are well separated 
and the $r_0$ values are explicitly shown on the curves. The (red) squares indicate
the results obtained with $C_1=C_0$ corresponding to the $SU(4)_W$ symmetry.}
\label{fig:fig3}
\end{figure}

As suggested from Efimov physics and pionless theory, we include now the (regularized) three-body contact
term, given in Eq.(\ref{eq:w30}). We solve the Schr\"odinger equation with two and three-body forces
for the different values of $r_0$ and $\beta$, with the strength $W_0$ fixed in such a way that in all
cases $B(^3{\rm H})=-8.48\;$ MeV. At this point predictions can be done for the doublet and quartet
$N-d$ scattering lengths, $^2a_{nd}\;\; ^4a_{nd}$, the corresponding low energies phases and the 
$^4$He energy, $B(^4{\rm He})$. As it is well known $^4a_{nd}$ is correlated
to the deuteron binding energy whereas $^2a_{nd}$ and $B(^4{\rm He})$
scale to some extent with $B(^3{\rm H})$. 
We have calculated the doublet and quartet $nd$ scattering lengths. 
In the first case
we obtain values of $^2a_{nd}\approx 0.3\div 0.5\;$fm, showing  some spread
in the results and, taking into account that this is a LO description, we consider these results 
compatible with the experimental value of $^2a_{nd}=0.65\pm0.01\;$fm~\cite{ndexp}.
For the quartet case the results are much stable, in all cases we obtain $^4a_{nd}\approx 6.35 \;$fm
in complete agreement with the experimental result. Results for the $^4$He energy are shown
in Fig.~\ref{fig:fig3} as a function of the regulator $\beta$ with and without the three-body interaction. 
Different values of the potential range $r_0$ have been considered, when the three-body term is included
the $r_0$ values are indicated in the box whereas it is explicitly shown on the curves when the three-body
term was not considered. In the first case we observe some spread around 
the experimental value of $-28.3\;$MeV, however the experimental energy is well described, within a $10\%$ accuracy.
Without the inclusion of the three-body LEC the results show an extremely large spread making the
pionless and chiral results arbitrarily different. 
In the figure the case $C_1=C_0$, verifying the Wigner spin-flavor symmetry is shown with solid (red) squares.
Moreover, for $B(^4{\rm He})$, 
the potential energy mean value of the two-body contact terms is around five times greater than 
the OPEP mean value which in turn is of the same order of magnitude of the three-body contact term
mean value. For all these observations we conclude that a LO three-body force is needed in both theories. 

What we want to stress here is the consequence of promoting the three-body contact
term to LO in the chiral expansion of the nuclear interaction. 
In the normal counting the three-body contact term considered here at LO, and
usually called $c_E$-term, appears at N2LO together with the $c_D$-term, a two-body contact term 
plus a pion exchange (see Ref.~\cite{epelbaum06}).  At N2LO two-pion exchange terms appear governed by the constants
$c_1$, $c_3$ and $c_4$. At this level of the chiral expansion the new LECs are 
$c_E$ and $c_D$. Numerical values were determined in the literature from a fit to observables 
as for example $B(^3{\rm H})$ and $^2a_{nd}$. In these fits usually the NN two-body potential was considered 
at N3LO. Consistency between the two- and three-body forces requires to construct
the three-body interaction at N3LO~\cite{n3lof1,n3lof2}. 
At this level however no new LECs appear. First applications of the complete N3LO nuclear interaction have 
addressed the description of the three-nucleon scattering, however some problems still remain in those
observables governed by $P$-waves~\cite{bochum}. This is an indication that spin structures appearing
at higher orders could be important parts of the three-nucleon interaction. 

Parallel to this analysis the contact three-body force at N4LO has been discussed in 
Ref.~\cite{girlanda}. In this study it was shown that the N4LO subleading contact interaction 
has 10 new LECs with the explicit expression given in that reference. The arguments given here 
to promote the $c_E$ term at LO entails a corresponding promotion of the N4LO subleading contact 
terms at N2LO, since they merely specify finer details of the 
short-distance interaction, of relative order $O(p^2)$ compared to the LO one. 
Accordingly the three-body force at N2LO will have, in addition to the $c_E$ appearing
at leading order, the contact-one-pion LEC $c_D$ 
plus the 10 LECs of the subleading terms. This mechanism will improve the convergence of the ChPT
since it increases the accuracy of the nuclear potential at N2LO which, considering only the
two-body part, has sufficient flexibility to describe NN phases with a $\chi^2$ per datum
$\approx 1$ up to $125$ MeV lab energy~\cite{nnopt}. It is thus sensible 
to investigate the three- and four-nucleon continuum using a N2LO interaction with two- and
three-body forces having in the three-body interaction 12 independent LECs. The capability of the
new terms to improve the theoretical description of spin observables has been recently 
investigated~\cite{girlanda2}. A more extended analysis is underway.

\section{Leading order formulation for shallow states}
In the following we analyze the consequences for the power counting of the additional  scale introduced by the shallow characteristic of the deuteron binding energy.
It is worth noticing that, while the underlying (approximate) chiral symmetry of strong interactions 
allows to explain the hierarchy of many-body forces, it has little to say about the 
large cancellation between the kinetic and potential energy produced in shallow states. 
On the basis of chiral counting both kinetic and potential energies are expected to be 
of order $O(p^2/\Lambda) \sim m_\pi^2/\Lambda \sim 20$~MeV, while the sum of the two is smaller by one 
order of magnitude in the deuteron (here $m_\pi$ is the pion mass and
$\Lambda\approx 1\;$GeV is the high momentum scale, see below). 
Whether this fine tuning is an accident or depends on some underlying  
physics is a question that need not concern us for the present discussion. 
In order to provide a rationale for the promotion of a three-body contact term to LO, 
we have to incorporate the small scale 
$\epsilon \sim 1/a$ associated with the large scattering length $a$ into the effective theory. It is well known 
that the Goldstone boson character of pions allows to estimate the size of interaction Lagrangians as 
\begin{equation}
{\cal L} = c_{\ell m n} \left( \frac{\bar N ... N}{f_\pi^2 \Lambda} \right)^\ell 
\left( \frac{\pi}{f_\pi}\right)^m \left(\frac{\partial_\mu, m_\pi}{\Lambda}\right)^n \Lambda^2 f_\pi^2,
\end{equation}
with dimensionless rescaled LECs $c_{\ell m n} \sim O(1)$ under the hypothesis of naturalness. This is 
called na\"ive dimensional analysis \cite{nda} (cfr. also \cite{friar}). $f_\pi$ is the scale of Goldstone 
bosons' fields, and sets the chiral symmetry breaking scale as $\Lambda \sim 4 \pi f_\pi$, with the factor 
$4\pi$ coming from momenta integrations. $\Lambda$ represents the mass scale of hadrons unprotected by chiral 
symmetry, e.g. $\Lambda \sim m_N$, the nucleon mass. 
The scaling of each factor  in the above formula can be simply obtained by 
applying it to the LO Lagrangian of interacting pions and nucleons \cite{friar}. The same can be done in a regime in 
which pions play no role, only nucleons are present, and a relevant scale is identified in $\epsilon$. 
The free nucleon Lagrangian is not enough to constrain the scaling of nucleon fields and derivatives. 
We also need an interaction term,
\begin{equation}
{\cal L}_0^{\slash{\!\!\!\pi}} = \bar N ( i \partial_\mu \gamma^\mu - m_N ) N -\frac{ D_0}{2} (\bar N N )^2,
\end{equation}
with the contact LEC $D_0$ related to the scattering length as 
$D_0 \approx 4 \pi a/m_N \sim 4\pi/m_N\epsilon \sim 1/(f_\pi \epsilon)$, and $f_\pi$ serving here only a mnemonic purpose.
This fixes uniquely the scaling of individual fields in the Lagrangian,
\begin{equation}
{\cal L} = c_{\ell m n} \left( \frac{\bar N ... N}{f_\pi \Lambda \epsilon} \right)^\ell  \left( \frac{\pi}{f_\pi}\right)^m
\left(\frac{\partial_\mu, m_\pi}{\Lambda}\right)^n \Lambda^2 f_\pi \epsilon.
\end{equation}
We then see that contact terms are naturally enhanced by the small scale $\epsilon$, e.g. a three-nucleon contact 
term would have a natural LEC $c_E$ of order $c_E \sim 1/(f_\pi^2 \Lambda \epsilon^2)$; in a combined small 
momenta and small $\epsilon$ expansion, it would have to be promoted to lower orders. 
The na\"ive dimensional analysis is recovered imposing that $\epsilon$ is natural, i.e.
$\epsilon \sim O(p) \sim f_\pi$. The two forms are not in contradiction since, in a matching between the
two regimes, there could be dimensionless ratios of the two scales $\epsilon/f_\pi$  restoring the proper scaling.
Thus we see that, allowing $\epsilon$ to be smaller than natural, pion-range interactions should actually be 
regarded as small compared to contact interactions, in agreement with the so-called KSW counting \cite{ksw}.
Therefore, if a three-body parameter is needed at LO in the pionless theory, the same is true in the chiral effective
theory. The increased importance of contact interactions in ChPT would also explain the crucial role 
they play for the construction of phenomenologically realistic interactions, both in the two- and three-nucleon sector, 
as well as in the nuclear interactions with external electroweak probes.
The proposed mechanism of promotion of contact terms, contrary to arguments based on the non-perturbative 
renormalization of the one-pion-exchange potential \cite{ntvk,birse,valderrama}  doesn't rely in any way on the role of 
pions. It just reflects the fact that the scattering length is larger than natural. In particular it is 
conceivable that, in a theory with no spontaneous chiral symmetry breaking, as in the case of QCD with one 
single flavor, gauge interactions (which are flavor-blind) could lead in principle to a similar scenario of 
quasi-universality.

\section{Conclusions} 
In the present work we have discussed two aspects of the two-, three-
and four-nucleon systems in a LO description. The main conclusion is that the $c_E$
term appearing in ChPT at N2LO has to be considered at LO due to the same situation
already analyzed in pionless theory (or Efimov physics) and related to the Thomas
collapse.
This promotion has the consequence of moving to lower orders subleading
contact terms, increasing the number of LEC's and allowing for a faster convergence of the
perturbative series. 
A second point regards the unusual large values of the two-nucleon
scattering lengths in singlet and triplet states.
It was shown that, considering the Wigner spin-flavor symmetry, the single LEC associated to the
central short-range potential, in combination with the OPEP is responsible for these large values. 
This observation reduces, to some extent, the amount of fine-tuning required to explain the
closeness of the nuclear interaction to the unitary limit.

{\sl Acknowledgement.} The authors thank R. Schiavilla, L.E. Marcucci and H.W. Hammer for useful discussions.

\end{document}